\documentclass[final,1p,times,twocolumn]{elsarticle}

\usepackage{lineno,hyperref}
\modulolinenumbers[5]

\journal{Journal of \LaTeX\ Templates}
\usepackage{epsfig}
\usepackage{amssymb}
\usepackage{amsmath}
\usepackage{bookmark}
\usepackage{rotating}
\usepackage{lscape}
\usepackage{longtable}
\usepackage{array}
\journal{Nuclear Physics B}

\newcommand{\ket}[1]{\left|#1\right\rangle}      

\def\l{\lambda}

\def\V={{{\bf\rm{V}}}}

\def\be{\begin{eqnarray}}
\def\ee{\end{eqnarray}}
\def\ba{\begin{array}}
\def\ea{\end{array}}
\def\no{\nonumber}

\begin{document}

\title{Bethe states  of the trigonometric $SU(3)$ spin chain with generic open boundaries}

\author[a,b]{Pei Sun,}
\author[a,b]{Zhirong Xin,}
\author[a,b]{Yi Qiao,}
\author[a,b]{Fakai Wen,}
\author[a,b]{Kun Hao,}
\author[c,d,e]{Junpeng Cao,\corref{cor1}}
\ead{junpengcao@iphy.ac.cn} \cortext[cor1]{Corresponding author}
\author[f]{Guang-Liang Li,}
\author[a,b]{Tao Yang }
\author[a,b]{Wen-Li Yang \corref{cor1}}
\ead{wlyang@nwu.edu.cn} 
\author[a,b]{Kangjie Shi}

\address[a]{Institute of Modern Physics, Northwest University,
Xian 710069, China}
\address[b]{Shaanxi Key Laboratory for
Theoretical Physics Frontiers,  Xian 710069, China}
\address[c]{Beijing National Laboratory for Condensed Matter
Physics, Institute of Physics, Chinese Academy of Sciences, Beijing
100190, China }
\address[d]{School of Physical Sciences,
University of Chinese Academy of Sciences, Beijing, China}
\address[e]{Collaborative Innovation Center of Quantum Matter,
Beijing,  China}
\address[f]{Department of Applied Physics, Xian
Jiaotong University, Xian 710049, China}

\date{\today}
\begin{abstract}
By combining the algebraic Bethe ansatz and the
off-diagonal Bethe ansatz, we investigate the trigonometric $SU(3)$
model with generic open boundaries. The eigenvalues of the transfer
matrix are given in terms of an inhomogeneous $T-Q$ relation, and
the corresponding eigenstates are expressed in terms of nested
Bethe-type eigenstates which have well-defined homogeneous limit. This exact
solution provides a basis for further analyzing the thermodynamic
properties and correlation functions of the anisotropic models associated with
higher rank algebras.
\end{abstract}


\maketitle

\section{Introduction}
\label{sec:introduction}
Since the work of Yang and Baxter's pioneering works~\cite{1cn Yang,2RJ Baxter}, the exactly solvable quantum
integrable systems~\cite{3Baxter 2} have attracted a great deal of interest. Especially the quantum
integrable models are very useful in nano-scale systems where alternative approaches involving mean field approximations or perturbations have failed~\cite{4CT Black,5J Dukelsky}. Moreover,
it also play very important role in statistical physics~\cite{6BM Mccoy}, low-dimensional consensed
matter physics~\cite{7T Giamarchi}, and even some mathematical areas such as quantum groups~\cite{8V Chari}.

By using the coordinate Bethe ansatz and the algebraic Bethe ansatz~\cite{10L.A. Takhtadzhan,11E.K. Sklyanin,12F.C. Alcaraz,13E.K. Sklyanin},
both eigenvalues and eigenstates of the integrable models with $U(1)$ symmetry can be given successfully.
For the integrable models without $U(1)$ symmetry (with generic boundary conditions), it has been proven that the off-diagonal Bethe ansatz is a powerful method~\cite{14J.Cao1,15J.Cao2,16J.Cao3,17J.Cao4,
18J.Cao5,19J.Cao6,20X.Zhang,21K.Hao} (For further information, we refer the reader to~\cite{22Y.Wang}). Based on the inhomogeneous $T-Q$ relation constructed via the
off-diagonal Bethe Ansatz, a systematic method~\cite{jia Zhang state1,jia zhang state2} for retrieving the Bethe-type
eigenstates (Bethe states) of integrable models without obvious reference state is also developed by employing certain orthogonal basis of the Hilbert space.
It should also be remark that the separation of variables (SoV) method~\cite{jia Niccoli1,
jia Niccoli2,jia Niccoli3}  and the modified algebraic Bethe ansatz
method \cite{1,2,3} were also
used to approach the eigenstates of several integrable models.

With the help of the off-diagonal Bethe ansatz~\cite{22Y.Wang}, the exact energy spectrum of the rational
$SU(3)$ model with unparallel boundary fields has been obtained~\cite{Guangliang Li}. However, the
eigenstates (or Bethe states) which have played important roles in  applications of the model are still missing. In this paper, the R-matrix is the trigonometric one associated with the $SU_q(3)$ algebra and the boundary reflection matrices are the most generic reflection matrices with non-zero off-diagonal elements. By combining  the nested algebraic Bethe ansatz and off-diagonal Bethe ansatz, we obtain the eigenvalues
which have well-defined homogeneous limit and the corresponding Bethe states of the transfer matrix of the model. Numerical results for the small size systems show that the
spectrum obtained by the nested Bethe ansatz equations (BAEs) is complete and the Bethe states are exactly correct.

The paper is organized as follows. In section~\ref{sec:The model}, the
$SU_{q}(3)$-invariant $R$-matrix and corresponding generic integral
non-diagonal boundary reflection matrices are introduced. In
section ~\ref{sec:Algebraic Bethe Ansatz}, we introduce the vacuum state of the system. In section 4, we construct the Bethe-type eigenstate of the transfer matrix.
In section 5, we introduce the
gauge transformation in order to construct the nested Bethe states.
In section~\ref{sec:Bethe states problem}, the nested Bethe states of the system are obtained. Section~\ref{sec:concluding remarks} gives some
discussions.

\section{The model}
\label{sec:The model} \setcounter{equation}{0}
Throughout this paper we adopt the standard notations:
for any matrix $A\in {\rm End}({\rm\bf V})$, $A_j$ is an
embedding operator in the tensor space ${\rm\bf
V}\otimes {\rm\bf V}\otimes\cdots$, which acts as $A$ on the $j$-th
space and as identity on the other factor spaces. For $R\in {\rm
End}({\rm\bf V}\otimes {\rm\bf V})$, $R_{ij}$ is an embedding
operator of $R$ in the tensor space, which acts as
identity on the factor spaces except for the $i$-th and $j$-th ones.

The $R$-matrix $R(u)\in {\rm End}({\rm\bf V}\otimes {\rm\bf V})$
used in this paper was first proposed by Perk and Shultz
\cite{361J.H.H. Perk} and further studied in
\cite{36J.H.H. Perk,37C.L. Schultz,38J.H.H. Perk,39V.V. Bazhanov,40M.Jimbo}. We now discuss the trigonometric one associated with the
$SU_q(3)$ algebra,
\begin{eqnarray}
  R_{12}(u)= \left(
                   \begin{array}{ccc|ccc|ccc}
                     a(u) &  &  &  &  &  &  &  &  \\
                      & b(u) &  & c(u) &  &  &  &  &  \\
                      &  & b(u) &  &  &  & c(u) &  &  \\
                     \hline
                      & d(u) &  & b(u) &  &  &  &  &  \\
                      &  &  &  & a(u) &  &  &  &  \\
                      &  &  &  &  & b(u) &  & c(u) &  \\
                     \hline
                      &  & d(u) &  &  &  & b(u) &  &  \\
                      &  &  &  &  & d(u) &  & b(u) &  \\
                      &  &  &  &  &  &  &  & a(u) \\
                   \end{array}
                 \right),
\end{eqnarray}
where the matrix elements are
\be
&&a(u)=\sinh(u+\eta), \quad
b(u)=\sinh(u), \nonumber \\
&&c(u)=e^{u}\sinh\eta, \quad d(u)=e^{-{u}}\sinh \eta.
\ee
The
$R$-matrix possesses the following   properties,
\be
\mbox{Initial condition}:&&\,R_{12}(0)= \sinh\eta P_{12},\nonumber\\
\mbox{ Unitarity relation}:&&\,R_{12}(u)R_{21}(-u)= \rho_1(u)\,\times {\rm id},\nonumber\\
\mbox{ Crossing Unitarity relation}:&& \,R_{12}^{t_1}(u)\,{\cal M}_1\,R_{21}^{t_1}(-u-3\eta)\,{\cal M}_1^{-1}=\rho_2(u)\,\times {\rm id},\nonumber \\
\mbox{ PT-symmetry}:&&\,R_{21}(u)=R^{t_1\,t_2}_{12}(u),\nonumber\\
\mbox{ Periodicity}: &&\, R_{12}(u+i\pi)=-R_{12}(u). \ee
Here $R_{21}(u)=P_{12}R_{12}(u)P_{12}$ with $P_{12}$ being the usual
permutation operator and $t_i$ denotes transposition in the $i$-th
space. The functions $\rho_1(u)$, $\rho_2(u)$ and the crossing
matrix ${\cal M}$ are given by
\be
\rho_1(u)&=&-\sinh({u}-\eta)\sinh({u}+\eta),\nonumber\\
\rho_2(u)&=&-\sinh({u})\sinh({u}+3\eta),\nonumber\\
{\cal
M}&=&\left(\begin{array}{ccc}e^{4\eta}&&\\&e^{2\eta}&\\&&1\end{array}\right).
\ee
The $R$-matrix satisfies the quantum Yang-Baxter equation (QYBE)
\be
R_{12}(u_1-u_2)R_{13}(u_1-u_3)R_{23}(u_2-u_3)
\!=\!R_{23}(u_2-u_3)R_{13}(u_1-u_3)R_{12}(u_1-u_2).
\label{QYBE}
\ee
Let us now introduce the reflection matrix $K^-(u)$ and the dual one
$K^+(u)$. The former satisfies the reflection equation (RE)
\begin{eqnarray}
\label{kfu RKRK}
 &&R_{12}(u_1-u_2)K^-_1(u_1)R_{21}(u_1+u_2)K^-_2(u_2)\nonumber\\
 &&=
K^-_2(u_2)R_{12}(u_1+u_2)K^-_1(u_1)R_{21}(u_1-u_2),\label{QREK1}
\end{eqnarray}
and the latter satisfies the dual RE
\be
&&R_{12}(u_2-u_1)K^+_1(u_1){\cal M}_1^{-1}R_{21}(-u_1-u_2-3\eta){\cal M}_1K^+_2(u_2)\no\\
&&= K^+_2(u_2){\cal M}_2^{-1}R_{12}(-u_1-u_2-3\eta){\cal M}_2K^+_1(u_1)R_{21}(u_2-u_1).
\label{QREK2}\ee
We now consider the generic non-diagonal $K$-matrices $K^-(u)$
\be
 K^-(u)\!&=&\!\left(\begin{array}{ccc}e^{u}\!\sinh(\zeta-u)\!+ce^{2u}\!\sinh (2u)\!&0&0\\
0&e^{u}\!\sinh(\zeta-u)\!&c_1\sinh (2u)\!\\
0&c_2\sinh (2u)\!&e^{-u}\!\sinh(\zeta+u)\!\end{array}\right)\nonumber \\
\!&=&\! \left(
      \begin{array}{ccc}
        k_{11}^{-} & 0 & 0 \\
        0 & k_{22}^{-} & k_{23}^{-} \\
        0 & k_{32}^{-} & k_{33}^{-} \\
      \end{array}
    \right),
\label{Kmatrixfu}
\ee
where the four boundary parameters $c$,$c_{1}$,$c_{2}$ and $\zeta$ satisfy a constraint
\be
c^2=c_1c_2+ce^{\zeta}. \no\ee
The dual non-diagonal reflection
matrix $K^+(u)$ is given by
\be
K^+(u)={\cal M}K^-(-u-3\eta/2)\left|_{(\zeta,c,c_1,c_2)\rightarrow
(\zeta',c',c_1',c_2')}\right.
=\left(
   \begin{array}{ccc}
        k_{11}^{+} & 0 & 0 \\
        0 & k_{22}^{+} & k_{23}^{+} \\
        0 & k_{32}^{+} & k_{33}^{+} \\
   \end{array}
 \right),\label{Kzheng matrix}
 \ee
with the constraint
\be
c'^2=c'_1c'_2+c'e^{\zeta'}. \no\ee
In order to show the intergrability of the system, we first
introduce the ``row-to-row" monodromy matrices $T_0(u)$
and $\hat{T}_0(u)$
\be
T_0(u)&=&R_{0N}(u-\theta_N)R_{0\,N-1}(u-\theta_{N-1})\cdots
R_{01}(u-\theta_1),\label{}\\
\hat{T}_0(u)&=&R_{10}(u+\theta_1)R_{20}(u+\theta_{2})\cdots
R_{N0}(u+\theta_N),
\ee
where $\{\theta_j, j=1, \cdots, N\}$ are the inhomogeneous
parameters and $N$ is the number
of sites. The one-row monodromy matrices are the $3\times 3$
matrices in the auxillary space $0$ and their elements act on the
quantum space ${\rm\bf V}^{\otimes N}$. For the system with open
boundaries, we need to define the double-row monodromy matrix
$\mathbb{T}_0(u)$
\be
\mathbb{T}_0(u)=T_0(u)K^-_0(u)\hat{T}_0(u)=\begin{pmatrix}
A(u) &  B_{1}(u) & B_{2}(u) \cr
C_{1}(u) & D_{11}(u) & D_{12}(u) \cr
C_{2}(u) & D_{21}(u) & D_{22}(u) \cr
\end{pmatrix}. \label{monodromy matrix}
\ee
which satisfies the similar relation as \eqref{kfu RKRK}
\begin{eqnarray}
 &&R_{12}(u_1-u_2)\mathbb{T}_1(u_1)R_{21}(u_1+u_2)\mathbb{T}_2(u_2)\nonumber\\
 &&=
\mathbb{T}_2(u_2)R_{12}(u_1+u_2)\mathbb{T}_1(u_1)R_{21}(u_1-u_2).\label{QRET1}
\end{eqnarray}
Then the transfer matrix of the system can be constructed as
\be
t(u)=tr_0\{K^+_0(u)\mathbb{T}_0(u)\}=\left[ k_{11}^{+}(u)A(u)+\sum_{i,j=1}^{2} k_{i+1,j+1}^{+}(u) D_{ji}(u)\right],
\label{transfer matrix}\ee
where $k_{ij}^{\pm}$ is the $K^{\pm}$ matrix element in the $i$th row and $j$th column.
Using Eq. QYBE (\ref{QYBE}), RE (\ref{QREK1}) and dual RE (\ref{QREK2}),
we obtain the commutativity of $t(u)$, which means $\left[t(u),t(v)\right]$=0.
The Hamiltonian of the model is constructed by taking the
derivative of the logarithm of the transfer matrix as \footnote{It was shown in \cite{Guangliang Li} that the transfer matrix has the property: $t(0)\varpropto {\rm id}$,
implying that the Hamiltonian given by (\ref{hamilton}) is actually proportion to $t'(0)$. }
\begin{eqnarray}
&&H=\sinh\eta \frac{\partial \ln t(u)}{\partial
u}|_{u=0,\{\theta_j\}=0}. \label{hamilton}
\end{eqnarray}

\section{Vacuum state}
\label{sec:Algebraic Bethe Ansatz} \setcounter{equation}{0}

The block-diagonal structure of the $K$-matrix (\ref{Kmatrixfu}) permits us to use the nested
algebraic Bethe ansatz to construct the associated Bethe states and obtain the eigenvalues
as follows.

We first introduce the reference state $\ket{\Psi_0}$ as
\begin{equation}
\label{reference state}
\ket{\Psi_{0}} = \bigotimes_{j=1}^{M} \ket{0}_{j}, \;\;\;\;\;\;\;
\ket{0}_{j}=\begin{pmatrix}
1 \cr 0 \cr 0 \cr
\end{pmatrix}.
\end{equation}
From the relations \eqref{monodromy matrix},\eqref{reference state}, the elements of matrix $\mathbb{T}_0(u)$  acting on the reference state $\ket{\Psi_0}$ give rise to
\begin{eqnarray}
A(u) \ket{\Psi_0} &=&k^{-}_{11}(u) a_{0}(u) \ket{\Psi_0}, \nonumber \\
D_{11} (u) \ket{\Psi_0} &=& \left \{ \frac{d(2u)k^{-}_{11}(u)}{a(2u)} a_{0}(u) + \left[ k^{-}_{22}(u) -  \frac{d(2u)}{a(2u)}{k^{-}_{11}(u)} \right] b_{0}(u) \right\} \ket{\Psi_0}, \nonumber \\
D_{22} (u) \ket{\Psi_0} &=& \left \{ \frac{d(2u)k^{-}_{11}(u)}{a(2u)} a_{0}(u) + \left[ k^{-}_{33}(u) -  \frac{d(2u)}{a(2u)}{k^{-}_{11}(u)} \right] b_{0}(u) \right\} \ket{\Psi_0}, \nonumber \\
D_{12} (u) \ket{\Psi_0} &=& k^{-}_{23}(u)  b_{0}(u) \ket{\Psi_0},
\qquad
D_{21} (u) \ket{\Psi_0} = k^{-}_{32}(u)  b_{0}(u) \ket{\Psi_0}, \nonumber \\
B_{i} (u) \ket{\Psi_0} &\neq& 0,
\qquad
C_{i} (u) \ket{\Psi_0} = 0,
\;\;\;\;\;\;\;\;\; i=1,2, \label{ref111erence state}
\end{eqnarray}
where
\begin{equation}
  b_{0}(u)=\prod_{j=1}^{N}\sinh(u+\theta_{j})\sinh(u-\theta_{j}),\quad\quad a_{0}(u)=b_{0}(u+\eta).
\end{equation}
It is easy to prove that the reference state \eqref{reference state} is an eigenstate of the transfer matrix
\begin{eqnarray}
&&t(u)\ket{\Psi_0} =\Lambda_{0}\ket{\Psi_0}, \nonumber \\
&& \Lambda_{0}=k_{11}^+(u)k_{11}^-(u)a_0(u)+\frac{d(2u)}{a(2u)}\sum_{i=1}^2 k_{i+1,i+1}^{+}(u)k_{11}^{-}(u)(a_0(u)-b_0(u))\nonumber\\
&&\qquad +\sum_{i,j=1}^2k_{i+1,j+1}^+(u)k_{j+1,i+1}^-(u)b_0(u).
\end{eqnarray}

\section{Bethe state}
\setcounter{equation}{0}

From \eqref{ref111erence state}, we see that the operators $B_{1}(u)$ and $B_{2}(u)$ acting on the reference
state give nonzero values, thus can be regarded as the creation
operators of the eigenstates of the system. Following the procedure
of the nested algebraic Bethe ansatz, the eigenstates of the
transfer matrix can be constructed  as

\begin{equation}
\label{bethe state}
\ket{\Psi}= B_{a_{M}}(u_{M}) B_{a_{M-1}}(u_{M-1}) \dots B_{a_{1}}(u_{1})
\mathcal{F}^{a_{1} a_{2} \dots a_{M}} \ket{\Psi_0},
\end{equation}
where we have used the convention that the repeated indices indicates the sum over the values $1$,$2$, and
$\{\mathcal{F}^{a_{1}\dots a_{M}}\}$ are some undetermined functions of the Bethe roots
$\{u_{j}\}$. In fact, the $\{\mathcal{F}^{a_{1}\dots a_{M}}\}$ are
the vector components of the nested Bethe states (see below \eqref{eigenvalue2}). As the transfer matrix \eqref{transfer matrix} acting on the assumed
states \eqref{bethe state}, we should exchange the positions
of the operators $A(u)$, $D_{ij}(u)$ and the operators
$B_{a_{j}}(u_j)$. Combining the reflection equation
\eqref{QRET1} and the Yang-Baxter equation \eqref{QYBE}, we can derive following
commutation relations
\begin{eqnarray}
A(u) B_{j}(\l) &\!=\!& \frac{a(\l - u)}{b(\l - u)} \frac{b(\l +u)}{a(\l + u)}
B_{j}(\l) A(u) \!-\! \frac{b(2\l )}{a(2\l )} \frac{c(u-\l)}{b(\l -u)} B_{j}(u) A(\l) \nonumber \\
&&-\frac{c(u+\l)}{a(u + \l)} B_{i}(u) \tilde{D}_{ij}(\l), \\
\tilde{D}_{ij} (u) B_{k}(\l) &\!=\!& \frac{r^{id}_{ef} (u +\l+\eta) r^{fg}_{kj} (u -\l)}{b(u+\l+\eta) b(u - \l)}
B_{d}(\l) \tilde{D}_{eg} (u) \!-\!\frac{r^{id}_{ej} (2u +\eta)c(\l-u)}{a(2u) b(u - \l)} B_{d}(u) \tilde{D}_{ek}(\l) \nonumber \\
&&+ \frac{b(2\l )d(u+\l)}{a(2\l)}\frac{r^{id}_{kj} (2u +\eta)}{a(2u) a(u + \l)} B_{d}(u) A(\l),
\label{222} \\
B_{i}(u) B_{j}(\l) &\!=\!& B_{k}(\l) B_{l}(u) \frac{r^{ij}_{lk}(u - \l)}{a(u - \l)},
\end{eqnarray}
where
\begin{eqnarray}
r(u) = \begin{pmatrix}
\sinh(\eta+u) & 0 & 0 & 0 \cr
0 & \sinh(u) & e^{u}\sinh\eta & 0 \cr
0 &  e^{-{u}}\sinh \eta & \sinh(u) & 0 \cr
0 & 0 & 0 & \sinh(\eta+u) \cr
\end{pmatrix},
\end{eqnarray}
and
\begin{equation}
\tilde{D}_{ij}(u)= D_{ij}(u) -\frac{\delta_{ij}d(2u)}{a(2u)} A(u).
\end{equation}
Acting the transfer matrix $t(u)$ on the assumed eigen-state \eqref{bethe state} and repeatedly using the
commutation relations, we obtain
\begin{eqnarray}
t(u)\ket{\Psi}&=&
\left[ \frac{d(2u)}{a(2u)} \sum_{i=2}^{3} k_{ii}^{+}(u)
+k_{11}^{+}(u) \right] k_{11}^{-}(u)
a_{0}(u) \prod_{i=1}^{M}
\frac{a(u_{i} -u)}{b(u_{i} - u)} \frac{b(u_{i} + u)}{a(u_{i} + u)} \ket{\Psi} \nonumber \\
&&+ b_{0}(u) \prod_{i=1}^{M}
\frac{1}{b(u-u_{i}) b(u+u_{i}+\eta)} \hat{\Lambda}(u,\{ u_{j} \})  \ket{\Psi}
 \nonumber \\
&&  + \mbox{unwanted terms}. \label{111}
\end{eqnarray}
The vector components $\{F^{a_{1}\dots a_{M}}\}$ in the Bethe state \eqref{bethe state} allow us to reconstruct the associated Bethe state
$\ket{\mathcal{F}}$, and $\hat{\Lambda}\left(u,\{u_{j} \}\right)$ in the \eqref{111} is the associated eigenvalue of the nested transfer matrix
$\hat{t}(u,\{u_{j}\})$
\begin{eqnarray}
\hat{t}(u,\{ u_{j} \}) &=&
e^{\eta}\frac{\sinh(2u)}{\sinh(2u+\eta)} \mbox{tr}_{\bar{0}}
\left[ \bar{K}^{+}_{\bar{0}} (u) r_{\bar{0} M}(u+u_{M}+\eta)
\dots r_{\bar{0}1}(u+u_{1}+\eta)\right.\nonumber\\
&&\left.\times\bar{K}^{-}_{\bar{0}} (u) r_{1 \bar{0}}(u-u_{1})
\dots r_{M \bar{0}}(u-u_{M}) \right], \label{transfer matrix2}
\end{eqnarray}
which appears due to the commutation relations \eqref{222} and
\begin{eqnarray}
\bar{K}^{+}(u)&=&e^{-\eta}\begin{pmatrix}
k^{+}_{22}(u) & k^{+}_{23}(u) \cr
k^{+}_{32}(u) & k^{+}_{33}(u) \cr
\end{pmatrix}\nonumber \\
&=&\left(
                        \begin{array}{cc}
                          e^{-(u+\frac{\eta}{2})}\sinh(\zeta'+\frac32 \eta+u) & e^{\eta}c'_{1}\sinh(-2u-3\eta) \\
                          e^{-\eta}c'_{2}\sinh(-2u-3\eta) & e^{u+\frac{\eta}{2}}\sinh(\zeta'-\frac32\eta-u) \\
                        \end{array}
                      \right),\label{kfumatrix2}
\\
\bar{K}^{-}(u)&=&\frac{\sinh(2u+\eta)}{\sinh(2u)}\begin{pmatrix}
k^{-}_{22}(u)\! -\! \frac{d(2u)k^{-}_{11}(u)}{a(2u)} & k^{-}_{23}(u) \cr
k^{-}_{32}(u) & k^{-}_{33}(u)\! - \!\frac{d(2u)k^{-}_{11}(u)}{a(2u)} \cr
\end{pmatrix}\nonumber \\
&=&\left(
                      \begin{array}{cc}
                       e^{u+\eta}\sinh(\zeta -u)\!-\!c\sinh(\eta) & c_{1}\sinh(2u+\eta)\\
                       c_{2}\sinh(2u+\eta)& e^{-u}\sinh(\zeta+u+\eta)-c\sinh(\eta) \\
                      \end{array}
                      \right).\label{kzhengmatrix2}
\end{eqnarray}
The eigen-equation
\begin{equation}
\hat{t}(u,\{ u_{j} \}) \ket{\mathcal{F}}=\hat{\Lambda}(u,\{u_{j} \}) \ket{\mathcal{F}},\nonumber
\end{equation}
will be determined in section~\ref{sec:Bethe states problem}. The eigenstate $ \ket{\mathcal{F}}$ can be decomposed in terms of
the basis $\{|a_1,a_{2}\cdots,a_M\rangle\,|a_i=1,2;i=1,\cdots,M\}$ of the $M$-fold tensor space of ${\mathcal{C}^2\bigotimes \mathcal{C}^2\bigotimes\cdots\bigotimes\mathcal{C}^2}$ as
\begin{equation}
\ket{\mathcal{F}}=\sum_{\{a_i=1\}}^2\mathcal{F}^{a_{1} a_{2} \dots a_{M}}\,|a_{1},a_{2},\cdots,a_M\rangle,\label{eigenvalue2}
\end{equation}
where the vector components $\{\mathcal{F}^{a_{1} a_{2} \dots a_{M}} \}$ will be determined by (\ref{bethe state2}) lately.
The third term in equation \eqref{111} represent the unwanted
terms. The unwanted terms should be zero gives rise to the $M$ Bethe roots must satisfy
the following Bethe ansatz equations (BAEs)
\begin{eqnarray}
  1&=&-K^{(1)}({u_{k}})\frac{\sinh(2u_{k}+3\eta)}{\sinh(2u_{k}+\eta)}\frac{a_0(u_{k})Q^{(1)}(u_{k}-\eta)}{b_0(u_{k})\hat{\Lambda}(u_{k} ,\{ u_{j} \})},\quad k=1\ldots M,\label{BAES01}
\end{eqnarray}
where
\begin{eqnarray}
Q^{(1)}(u)&\!=\!&\prod_{i=1}^{M}\sinh(u-u_i)\sinh(u+u_i+\eta),\\
K^{(1)}(u)&\!=\!&(e^{u}\sinh(\zeta-u)+ce^{2u}\sinh2u)
\no\\[4pt]
&&\times (e^{-u+\frac {3\eta}
2}\sinh(\zeta'+u+\frac{\eta}{2})-c'e^{-2u+\eta}
\sinh(2u+\eta).
\end{eqnarray}

\section{Gauge transformation}
\setcounter{equation}{0}

\setcounter{equation}{0}
In the previous section, we have reduced constructing eigenstates of the
original transfer matrix $t(u)$ defined by \eqref{transfer matrix} into the Bethe states problem
of the nested transfer matrix $\hat{t}(u ,\{ u_{j} \})$ given by \eqref{transfer matrix2}.
Now we construct the eigenstate
$\ket{\mathcal{F}}$ of the nested transfer matrix $\hat{t}(u ,\{ u_{j} \})$.
Considering the
reflection matrices \eqref{kfumatrix2} and \eqref{kzhengmatrix2} have the
off-diagonal elements, the  $\hat{t}(u ,\{ u_{j} \})$ doesnot have the obvious
reference state, which means the analogy of construction of $\Psi_0$ is
invalid. Thanks to the works \cite{22Y.Wang}, we can solve the nested Bethe states problem \eqref{eigenvalue2} as follows.

For simplicity, we take the notation $\lambda=u+\frac12\eta$ and
$\lambda_{j}=u_{j}+\frac12\eta$. We recognize the $\hat{t}(u ,\{
u_{j} \})$ as the transfer matrix of the open spin-$1/2$ chain
of length $M$ with non-diagonal boundary terms.

We first introduce some parameterization:
\begin{eqnarray}
  e^{\theta_{-}} &=& 2p c_{1}, \qquad e^{\theta_{+}} = 2q c_{1}',\label{state relation1} \nonumber \\
  2\sinh^{2}(\alpha_{-}) &=&
  -\frac{4+\bar{\xi_{2}}^{2}-\bar{\xi_{1}}^{2}}{4}
  +\sqrt{\frac{(4+\bar{\xi_{2}}^{2}-\bar{\xi_{1}}^{2})^{2}}{16}+\bar{\xi_{1}}^{2}}
  ,\nonumber\\
  2\sinh^{2}(\beta_{-}) &=&
  -\frac{4+\bar{\xi_{1}}^{2}-\bar{\xi_{2}}^{2}}{4}+
  \sqrt{\frac{(4+\bar{\xi_{1}}^{2}-\bar{\xi_{2}}^{2})^{2}}{16}+\bar{\xi_{2}^{2}}},\nonumber\\
  \sinh^{2}(-\beta_{+}) &=&-\frac{1-q^{2}}{2}+\sqrt{\frac{(1-q^{2})^{2}}{4}+q^{2} \cosh^{2}\zeta' },\nonumber\\
  \sinh^{2}(-\alpha_{+}) &=&
  \frac{q^{2} \sinh^{2} \zeta'}{1+\sinh^{2}(-\beta_{+})}\label{state relation2},
\end{eqnarray}
where
\begin{eqnarray}
&&  1=4p^{2}c_{1}c_{2}, \qquad   1=4q^{2}c_{1}'c_{2}', \nonumber\\
&&  \bar{\xi_{1}} = p(e^{\zeta+\eta}-2c\sinh(\eta)-e^{-\zeta}), \qquad
  \bar{\xi_{2}} = p(-e^{\zeta+\eta}+2c\sinh(\eta)-e^{-\zeta}).
\end{eqnarray}
Then the non-diagonal reflection matrices \eqref{kfumatrix2} and \eqref{kzhengmatrix2} read
\begin{eqnarray}
  \tilde{K^{-}}(\lambda)=\frac{1}{2p}
   \left(
  \begin{array}{cc}
  \bar{k}^{-}_{11}(\lambda) & \bar{k}^{-}_{12}(\lambda)\\[4pt]
  \bar{k}^{-}_{21}(\lambda)& \bar{k}^{-}_{22}(\lambda) \\
  \end{array}
  \right),\qquad\quad
  \tilde{K^{+}}(\lambda) =\frac{1}{2q}
  \left(
  \begin{array}{cc}
  \bar{k}^{+}_{11}(\lambda) & \bar{k}^{+}_{12}(\lambda)\\[4pt]
  \bar{k}^{+}_{21}(\lambda)& \bar{k}^{+}_{22}(\lambda) \\
  \end{array}
  \right),\label{state kmatrix2-2}
\end{eqnarray}
where
\begin{eqnarray}
  \bar{k}^{-}_{11}(\lambda) &=& 2[\sinh(\alpha_{-})\cosh(\beta_{-})\cosh(\lambda)+\cosh(\alpha_{-})\sinh(\beta_{-})\sinh(\lambda)],  \nonumber\\
  \bar{k}^{-}_{22}(\lambda) &=&
  2[\sinh(\alpha_{-})\cosh(\beta_{-})\cosh(\lambda)-\cosh(\alpha_{-})\sinh(\beta_{-})\sinh(\lambda)],
   \nonumber\\
  \bar{k}^{-}_{12}(\lambda) &=& e^{\theta_{-}}\sinh(2\lambda),\quad
  \bar{k}^{-}_{21}(\lambda) = e^{-\theta_{-}}\sinh(2\lambda), \nonumber\\
  \tilde{K}^{+}(\lambda)&=&\tilde{K}^{-}(-\lambda-\eta)|_{(\alpha_{-},\beta_{-},\theta_{-},p)\rightarrow
  (-\alpha_{+},-\beta_{+},\theta_{+},q)}.
\end{eqnarray}
By using the above relations, the nested transfer matrix can be expressed as
\begin{eqnarray}
 \hat{t}(\lambda ,\{ \lambda_{j}\}) &=&  e^{\eta}\frac{\sinh(2\lambda-\eta)}{\sinh(2\lambda)}
   \sigma_{1}(-\lambda_{1})\cdots \sigma_{M}(-\lambda_{M}) Tr_{0}(\tilde{K}_{0}^{+}(\lambda)
   \tilde{T}_{0}(\lambda)\tilde{K}_{0}^{-}(\lambda)\hat{\tilde{T}}_{0}^{+}(\lambda))\nonumber\\
   &&\times\sigma_{1}(\lambda_{1})\cdots \sigma_{M}(\lambda_{M}),
\end{eqnarray}
where
\begin{eqnarray}
       \sigma(\lambda) &=& \left(
                             \begin{array}{cc}
                               e^{\frac{1}{2}\lambda} &  \\
                                & e^{-\frac{1}{2}\lambda} \\
                             \end{array}
                           \right),
        \\
\tilde{T}_{\bar{0}}(\lambda,\{{\lambda}_j\})
&=& \tilde{R}_{\bar{0} M}(\lambda+\lambda_{M})
\dots \tilde{R}_{\bar{0}1}(\lambda+\lambda_{1}),\nonumber\\
\widehat{\tilde{T}}_{\bar{0}}(\lambda,\{{\lambda}_j\})
&=& \tilde{R}_{1 \bar{0}}(\lambda-\lambda_{1})
\dots \tilde{R}_{M \bar{0}}(\lambda-\lambda_{M}),\\
       \tilde{R}_{0j}(\lambda) &=& \left(
                                     \begin{array}{cccc}
                                       \sinh(\lambda+\eta) & 0 & 0 & 0 \\
                                        0 & \sinh(\lambda) & \sinh(\eta) & 0 \\
                                        0 & \sinh(\eta) & \sinh(\lambda) & 0 \\
                                        0 & 0 & 0 & \sinh(\lambda+\eta) \\
                                     \end{array}
                                   \right).
     \end{eqnarray}
Now, we introduce the gauge transformations
\begin{eqnarray}
 {X}_{m}(\lambda)&=& \left(
                       \begin{array}{c}
                         e^{-[\lambda+(\alpha+m) \eta]} \\
                         1 \\
                       \end{array}
                     \right)
 ,\quad
  Y_{m}(\lambda)=
  \left(
    \begin{array}{cc}
      e^{-[\lambda+(\alpha-m)\eta]} \\  1
    \end{array}
  \right),
 \nonumber \\
  \bar{X}_{m}(\lambda)&=& \frac{e^{\lambda+\alpha \eta}}{2 \sinh(m\eta)}
  \left(
    \begin{array}{cc}
      1 & -e^{-[\lambda+(\alpha+m)\eta]} \\
    \end{array}
  \right),\nonumber \\
  \hat{X}_{m}(\lambda)&=& \frac{e^{-\eta}\sinh((m+2)\eta)}{\sinh((m+1)\eta)}
  \left(
    \begin{array}{c}
      e^{-[\lambda+(\alpha+m)\eta]} \\
      1 \\
    \end{array}
  \right),\nonumber \\
  \bar{Y}_{m}(\lambda)&= &\frac{e^{\lambda+\alpha \eta}}{2 \sinh(m\eta)}
  \left(
    \begin{array}{cc}
      -1 & e^{-[\lambda+(\alpha-m)\eta]} \\
    \end{array}
  \right),\nonumber  \\
    \alpha \eta &=& -\theta_{+}+\eta+i\frac{\pi}{2}, \qquad
  m \eta = \alpha_{+}+\beta_{+}-i\frac{\pi}{2}, \nonumber  \\
    (\bar{m}+M)\eta &=& -\theta_{-}-\alpha \eta+\alpha_{-}+\beta_{-}+i\pi,
\end{eqnarray}
By using these gauge transformation, the nested transfer matrix can also be written as
\begin{eqnarray}
  \hat{t}(\lambda ,\{ \lambda_{j}\}) &=& \bar{Y}_m(-\lambda)\tilde{K}^+(\lambda)X_m(\lambda)\bar{A}_m(\lambda)
  +\bar{X}_{m}(-\lambda)\tilde{K}^+(\lambda)Y_m(\lambda)\bar{D}_m(\lambda),
\end{eqnarray}
where
\begin{eqnarray}
  \bar{A}_{m}(\lambda) &=& \bar{Y}_{m}(\lambda)\tilde{T}(\lambda)\tilde{K}^{-}(\lambda)
  \hat{\tilde{T}}(\lambda)\hat{X}_{m-2}(-\lambda),\nonumber \\
  \bar{D}_{m}(\lambda) &=& \bar{X}_{m}(\lambda)\tilde{T}(\lambda)\tilde{K}^{-}(\lambda)
  \hat{\tilde{T}}(\lambda)\hat{Y}_{m+2}(-\lambda).
\end{eqnarray}

\section{Nested Bethe states}
\label{sec:Bethe states problem}
\setcounter{equation}{0}
Now, we construct the eigenstate $\ket{\mathcal{F}}$ in \eqref{eigenvalue2} as \cite{jia Zhang state1}
\begin{eqnarray}
\label{bethe state2}
\ket{\mathcal{F}}&=& \sigma_{1}(-\lambda_{1})\cdots \sigma_{M}(-\lambda_{M})|w_{1},\cdots, w_{M};m;\bar{m}\rangle \nonumber \\
&=& \sigma_{1}(-\lambda_{1})\cdots \sigma_{M}(-\lambda_{M}) \bar{C}_{m}(w_{1})\bar{C}_{m+2}(w_{2}) \cdots \bar{C}_{m+2(M-1)}(w_{M})|\bar{m}\rangle,\nonumber\\
&=&\sum_{\{a_i=1\}}^2\mathcal{F}^{a_{1} a_{2} \dots a_{M}}\,|a_1,a_2,\cdots,a_M\rangle,
\end{eqnarray}
where the $\bar{C}_m(\lambda)$ matrix and the reference state are defined as
\begin{eqnarray}
\bar{C}_{m}(\lambda) &=& \bar{X}_{m}(\lambda)\tilde{T}(\lambda)\tilde{K}^{-}(\lambda)
  \hat{\tilde{T}}(\lambda)\hat{X}_{m}(-\lambda),\nonumber \\
|\bar{m}\rangle &=&\bigotimes_{n=1}^{M}e^{-[-\lambda_{n}+(\bar{m}+M-n+1+\alpha)\eta]}|\uparrow\rangle_{n}+|\downarrow\rangle_{n}.
\end{eqnarray}


From the reflection equation, we can obtain following commutation relations
\begin{eqnarray}
  \bar{D}_{m-2}(u_2)\bar{C}_{m-2}(u_1) &=&
  \frac{\sinh(u_1-u_2+\eta)\sinh(u_1+u_2)}{\sinh(u_1+u_2+\eta)\sinh(u_1-u_2)}\bar{C}_{m-2}(u_1)\bar{D}_{m}(u_2)\nonumber\\
  &-&\frac{\sinh(m\eta-u_1+u_2)\sinh(u_1+u_2)\sinh(\eta)}{\sinh(m\eta)\sinh(u_1-u_2)\sinh(u_1+u_2+\eta)}
  \bar{C}_{m-2}(u_2)\bar{D}_{m}(u_1)\nonumber\\
  &-&\frac{\sinh(m\eta+u_1+u_2)\sinh(\eta)}{\sinh(m\eta)\sinh(u_1+u_2+\eta)}\bar{C}_{m-2}(u_2)\bar{A}_{m}(u_1),\\
  \bar{A}_m(u_2)\bar{C}_m(u_1)&=&
  \frac{\sinh(u_1-u_2)\sinh(u_1+u_2+\eta)}{\sinh(u_1+u_2)\sinh(u_1-u_2+\eta)}\bar{C}_m(u_1)\bar{A}_{m+2}(u_2)
  \nonumber\\
  &+& \frac{\sinh(\eta)\sinh(u_1-u_2)\sinh(m\eta-u_2-u_1)}{\sinh(m\eta)\sinh(u_1+u_2)\sinh(u_1-u_2+\eta)}
  \bar{D}_m(u_1)\bar{C}_m(u_2)\nonumber\\
  &+&
  \frac{\sinh(m\eta+u_1-u_2)\sinh(\eta)}{\sinh(u_1-u_2+\eta)\sinh(m\eta)}\bar{A}_m(u_1)\bar{C}_m(u_2).
\end{eqnarray}
Acting the transfer matrix $\hat{t}(\lambda)$ on the state $\ket{\mathcal{F}}$ and repeatedly using the
commutation relations, we obtain
\begin{equation}
\hat{t}(\lambda,\{ \lambda_{j} \}) \ket{\mathcal{F}} =
\hat{\Lambda}(\lambda,\{\lambda_{j} \}) \ket{\mathcal{F}}+\mbox{unwanted terms}.
\label{eigenvalue2-sun}
\end{equation}
The first term in \eqref{eigenvalue2-sun} gives the eigenvalue $\hat{\Lambda}(\lambda)$ \cite{jia Zhang state1}
\begin{eqnarray}
\hat{\Lambda}(\lambda)&=&
\frac{\sinh(2\lambda-\eta)\sinh(2\lambda+2\eta)}{\sinh(2\lambda)\sinh(2\lambda+\eta)} K^{(2)}(\lambda)\bar{a}(\lambda)
\frac{Q^{(2)}(\lambda-\eta)}{Q^{(2)}(\lambda)}
\nonumber \\
&&\, +\frac{\sinh(2\lambda-\eta)}{\sinh(2\lambda+\eta)} K^{(3)}(\lambda)\bar{d}(\lambda)
\frac{Q^{(2)}(\lambda+\eta)}{Q^{(2)}(\lambda)}
\nonumber\\
&&\, +\,\sinh(2\lambda-\eta)\sinh(2\lambda+2\eta)\bar{a}(\lambda)\bar{a}(-\lambda-\eta)
\frac{h}{Q^{(2)}(\lambda)},\label{eigenvalue new}
\end{eqnarray}
\noindent
where
\be
\bar{a}(\lambda)&\!=\!&
\prod_{j=1}^M\sinh(\lambda+{\lambda}_j+\eta)\sinh(\lambda-{\lambda}_j+\eta),
\nonumber \\[-0.8mm]
\bar{d}(\lambda)&\!=\!&
\prod_{j=1}^M\sinh(\lambda-{\lambda}_j)\sinh(\lambda+{\lambda}_j),\nonumber \\
K^{(2)}(\lambda)&\!=\!& -\frac{1}{pq}e^{\eta}\sinh(\lambda-\alpha_{-})\cosh(\lambda-\beta_{-})\sinh(\lambda-\alpha_{+})
\cosh(\lambda-\beta_{+}),\nonumber \\
K^{(3)}(\lambda)&\!=\!&-\frac{1}{pq}e^{\eta}\sinh(-\lambda-\eta-\alpha_{-})\cosh(-\lambda-\eta-\beta_{-})\nonumber\\
&&\times \sinh(-\lambda-\eta-\alpha_{+})
\cosh(-\lambda-\eta-\beta_{+}),\nonumber \\
h&\!=\!& \frac{1}{2pq}e^{\eta}\cosh\left[(M+1)\eta+\alpha_{-}+\beta_{-}+\alpha_{+}+\beta_{+}\right]-\frac{1}{2pq}e^{\eta}\cosh(\theta_{-}-\theta_{+}),\nonumber \\
 Q^{(2)}(\lambda)&\!=\!&\prod_{j=1}^M \sinh(\lambda-w_j)\sinh(\lambda+w_j+\eta)\nonumber\\[-0.8mm]
&=&\prod_{j=1}^M \sinh(\lambda-g_j-\frac{1}{2}\eta)\sinh(\lambda+g_j+\frac{3}{2}\eta).
\ee
The second term in \eqref{eigenvalue2-sun} should be zero, which ensure the $\ket{\mathcal{F}} $ is the eigenstate of the nested transfer matrix \cite{jia Zhang state1,1,2,3}. This requires that
the $M$ Bethe roots $g_{l}$ must satisfy the BAEs
\begin{eqnarray}
1&+&\frac{\sinh(2g_l+{3}\eta)}{\sinh(2g_l+\eta)}\frac{K^{(2)}(g_l+\frac{1}{2}\eta)}{K^{(3)}(g_l+\frac{1}{2}\eta)}
\frac{Q^{(1)}(g_l+\eta)Q^{(2)}(g_l-\frac{1}{2}\eta)}{Q^{(1)}(g_l)Q^{(2)}(g_l+\frac{3}{2}\eta)}\no\\[6pt]
&=&-h\,{\sinh(2g_l+2\eta)\sinh(2g_l+3\eta)}
\frac{Q^{(1)}(g_l+\eta)}
{K^{(3)}(g_l+\frac{1}{2}\eta)Q^{(2)}(g_l+\frac{3}{2}\eta)},\nonumber\\
&& l=1,\ldots, M.
\label{BAES02}
\end{eqnarray}

\begin{table}[t]
\centering 
\begin{scriptsize}
\begin{tabular}{|c ccc|c|c|c|c|} \hline
$u_{1}$ & $u_{2}$ & $g_{1}$ & $g_{2}$ & $E_{n}$ & $n$  \\\hline
$0.1845 - 0.0000i $ & $0.2420+0.0000i$ &  $0.1783 - 0.0000i$ & $0.2603 + 0.0000i$ & $5.3807982858$ & $1$ \\
$-0.3628+0.0000i$ & $-$ & $0.1645 + 0.0000i$ & $-$ & $4.8827952486$ & $2$\\
$-$ & $-$ & $-$ & $-$ & $3.5453692295$ & $3$\\
$0.1647 + 0.0000i$ & $-0.1000 - 0.1706i$ &  $0.1699 + 0.0000i$ & $-0.2000 + 0.1909i$ & $2.7734720648$ & $4$\\
$-0.1000 + 0.2409i$ & $-$ & $-0.2000 + 0.2781i$ & $-$ & $2.3345140250$ & $5$\\
$-0.2042 + 0.0911i$ & $-0.2042 - 0.0911i$ & $-0.0381 - 0.0661i$ & $-0.0381 + 0.0661i$ & $0.6052190547$ & $6$\\
$-0.1000 - 0.1100i$ & $-$ & $0.0103 - 0.0000i$ & $-$ & $-0.1238250060$ & $7$\\
$-0.1000 + 0.0436i$ & $0.1458 - 0.0000i$ & $0.1603 - 0.0000i$ & $-0.2000 + 0.0137i$ & $-1.6800801819$ & $8$\\
$-0.1000+0.0566i$ & $-$ & $-0.2000 - 0.0351i$ & $-$ & $-2.5834417305$ & $9$\\
\hline \end{tabular}
\end{scriptsize}
\caption{\label{table1} Solutions of BAEs
\eqref{BAES1} and \eqref{BAES2} where $N=2$ with the
parameters $\eta=0.2, \zeta=0.1, c=1, c_{1}=-0.3,
\zeta'=-0.4, c'=-0.3, c_{1}'=-0.7, \alpha_-=0.8940, \alpha_+= 0.2704, \beta_-=-0.0321-1.5708i, \beta_+=0.8573, \theta_-=-0.0779-1.5708i, \theta_+=0.2604+3.1416i, p=1.5418i$ and $q=0.9267$ for  the case of $E_{n}$ is
the corresponding eigenenergy. The energy $E_{n}$ calculated from
\eqref{energy} is the same as that from the exact diagonalization
of the Hamiltonian \eqref{hamilton}.}
\end{table}
\begin{landscape}
\begin{table}[t]
\centering 
\begin{scriptsize}
\begin{tabular}{|c|c|c|c|c|c|c|c|}
\hline $u_1$ & $u_2$ & $u_3$ & $g_1$ & $g_2$ & $g_3$ & $E$ & $n$  \\ \hline
  $0.3661+0.0946i$ & $0.3661-0.0946i$ &  $0.1700+0.0000i$  & $0.3703+0.0878i$ &  $0.3703-0.0878i$ & $0.1701+0.0000i$ & $7.466014$ & $1$  \\
  $0.1709 - 0.0000i$  &  $0.3282 + 0.0000i$ & $-$ & $0.1708 - 0.0000i$  &  $0.3322 - 0.0000i$ & $-$ & $7.267876$ & $2$  \\
  $0.1688+0.0000i$  & $-$ & $-$ &  $0.1692 + 0.0000i$  & $-$ & $-$ &  $6.852899$ & $3$  \\
  $0.1721+0.0000i$ &  $-0.1000+0.2269i$ & $0.3079-0.0000i$ & $0.3177-0.0000i$ & $-0.2000+0.2548i$ & $0.1715+0.0000i$ & $5.968136$ & $4$  \\
  $0.1695 - 0.0000i$  &  $-0.1000 - 0.2610i$ & $-$ & $-0.2000 - 0.2956i$  &  $0.1703 - 0.0000i$ & $-$ & $5.786500$ & $5$  \\
  $-$ & $-$ & $-$ & $-$ & $-$ & $-$ & $5.585503$ & $6$ \\
  $-0.1000 - 0.3413i$  & $-$ & $-$ &  $-0.2000 - 0.3990i$ & $-$ & $-$ & $4.921459$ & $7$  \\
  $-0.2309 + 0.2518i$  &  $0.0309 + 0.2518i$ & $-$ & $-0.3877 + 0.2796i$  &  $-0.0123 + 0.2796i$ & $-$ & $4.324839$ & $8$  \\
  $0.0085 - 0.1597i$ &  $0.0085 + 0.1597i$  & $0.2049 - 0.0000i$ & $-0.0383 - 0.1605i$ & $-0.0383 + 0.1605i$ & $0.2088 - 0.0000i$ & $4.308177$ & $9$  \\
  $-0.1000 - 0.1375i$  &  $0.1534 + 0.0000i$ & $-$ & $0.1517 + 0.0000i$  &  $0.0430 - 0.0000i$ & $-$ & $4.231025$ & $10$  \\
  $0.1000 + 0.1929i$  & $-$ & $-$ &  $-0.0725 + 0.0000i$ & $-$ & $-$ & $3.852817$ & $11$  \\
  $-0.2027 - 0.1795i$  &  $0.0027 - 0.1795i$ & $-$ & $-0.0618 - 0.0000i$  &  $-0.2000 - 0.0720i$ & $-$ & $3.373027$ & $12$  \\
  $0.0869 - 0.0834i$ & $0.0869 + 0.0834i$ & $-0.1000 + 0.0900i$ & $0.0813 - 0.0768i$ & $0.0813 + 0.0768i$ & $-0.2000 + 0.0533i$ & $3.291098$ & $13$  \\
  $0.2769+0.0000i$ & $-0.1000-0.0857i$ &  $0.1778-0.0000i$ & $0.2920+0.0000i$ & $0.1751-0.0000i$ & $-0.2000-0.0815i$ & $2.673798$ & $14$  \\
  $0.1703 + 0.0000i$  &  $-0.1000 - 0.0902i$ & $-$ & $-0.2000 - 0.0879i$  &  $0.1721 + 0.0000i$ & $-$ & $2.370817$ & $15$  \\
  $-0.1000 - 0.1079i$  & $-$ & $-$ &  $-0.2000 - 0.1137i$ & $-$ & $-$ & $1.836570$ & $16$  \\
  $0.2439-0.0000i$ & $-0.1364+0.0000i$ & $0.0637+0.0000i$ & $0.2574-0.0000i$ & $0.0626+0.0000i$ & $-0.1286+0.0000i$ & $1.764730$ & $17$  \\
  $0.0348-0.0000i$  &  $-0.1652-0.0000i$ & $-$ & $0.0458+0.0000i$  &  $-0.1181+0.0000i$ & $-$ & $1.377880$ & $18$  \\
  $0.1034 - 0.0000i$  &  $-0.1000 + 0.0314i$ & $-$ & $0.1009 + 0.0429i$  &  $0.1009 - 0.0429i$ & $-$ & $0.769410$ & $19$  \\
  $-0.1000-0.0336i$ & $0.1851-0.0000i$ & $-0.4687-0.0000i$ & $-0.1582-0.0000i$ &  $0.2863+0.0000i$ & $0.1797+0.0000i$ & $0.038944$ & $20$  \\
  $0.1703 - 0.0000i$  &  $-0.1000 - 0.0342i$ & $-$ & $-0.1601 - 0.0000i$  &  $0.1733 - 0.0000i$ & $-$ & $-0.400469$ & $21$  \\
  $-0.1000 - 0.0391i$  & $-$ & $-$ &  $-0.1672 - 0.0000i$ & $-$ & $-$ & $-1.427163$ & $22$  \\
  $-0.1000+0.2375i$  &  $-0.1000-0.0617i$ & $-$ & $0.0741 + 0.0000i$  &  $-0.2000 + 0.3068i$ & $-$ & $-1.516419$ & $23$  \\
  $-0.1000+0.1444i$ &  $-0.1000+0.0414i$ & $0.2350-0.0000i$ & $0.2433-0.0000i$ & $-0.2845+0.0673i$ & $-0.1155+0.0673i$ & $-3.191544$ & $24$  \\
  $-0.1000-0.0435i$ & $-0.1000+0.1749i$ & $-$ & $-0.2648-0.0869i$ & $-0.1352-0.0869i$ & $-$ & $-3.226123$ & $25$  \\
  $-0.1000+0.0601i$  & $-$ & $-$ &  $0.0316+0.0000i$ & $-$ & $-$ & $-0.355535$ & $26$  \\
  $-0.1000+0.0418i$ & $-0.1000+0.1349i$ & $-$ & $0.0061+0.0000i$ & $-0.2705-0.0000i$ & $-$ &
  $-4.178601$ & $27$ \\
  \hline \end{tabular} \caption{\label{table2}Solutions of BAEs
\eqref{BAES1} and \eqref{BAES2} where $N=3$ with the
parameters $\eta=0.2, \zeta=0.1, c=1, c_{1}=-0.3,
\zeta'=-0.4, c'=-0.3, c_{1}'=-0.7, \alpha_-=0.8940, \alpha_+= 0.2704, \beta_-=-0.0321-1.5708i, \beta_+=0.8573, \theta_-=-0.0779-1.5708i, \theta_+=0.2604+3.1416i, p=1.5418i$ and $q=0.9267$ for  the case of $E_{n}$ is
the corresponding eigenenergy. The energy $E_{n}$ calculated from
\eqref{energy} is the same as that from the exact diagonalization
of the Hamiltonian \eqref{hamilton}.}
\end{scriptsize}
\end{table}
\end{landscape}

\begin{table}[htbp]
\centering 
\begin{scriptsize}
\begin{tabular}{|c|c|c|c|c|} \hline
$ F^{a_{2}a_{1}}$ & $B_{a_2}(u_{2})B_{a_{1}}(u_1)F^{a_{2}a_{1}}\ket{\Psi_0}$ & $E_{n}$ & $H\ket{\Psi_n}-E_n\ket{\Psi_n}$ & $n$ \\\hline
$\left(
  \begin{array}{c}
-0.5508 - 0.0000i\\
   1.0607 + 0.0000i\\
   1.3430 + 0.0000i\\
  -2.5973 - 0.0000i\\
  \end{array}
\right)$ &
$\left(
  \begin{array}{c}
        0\\
   0\\
   0\\
   0\\
  -1.0388 - 0.0000i\\
   2.0672 + 0.0000i\\
   0.0000 + 0.0000i\\
   2.5979 + 0.0000i\\
  -5.1455 - 0.0000i\\
  \end{array}
\right)$ &
$5.3807982858$ & $ 10^{-13}$ & $1$\\\hline
$\left(
  \begin{array}{c}
   0.0000 + 6.9054i\\
   0.0000 -45.1930i\\
  \end{array}
\right)$ &

$\left(
  \begin{array}{c}
  0\\
  -0.0000 + 0.6257i\\
   0.0000 - 1.2568i\\
  -0.0000 + 0.2872i\\
        0\\
        0\\
   0.0000 - 0.6281i\\
        0\\
        0\\
  \end{array}
\right)$ &
$4.8827952486$ & $ 10^{-13}$ & $2$\\\hline
$-$ &
$\left(
  \begin{array}{c}
        1  \\
        0  \\
        0  \\
        0  \\
        0\\
        0\\
        0  \\
        0\\
        0\\
  \end{array}
\right)$ & $3.5453692295$ & $0$ & $3$\\\hline
$\left(
  \begin{array}{c}
   0.3696 + 0.4676i\\
  -0.4734 - 0.2248i\\
  -0.4753 - 0.7129i\\
   0.4996 + 0.2026i\\
  \end{array}
\right)$ &

$\left(
  \begin{array}{c}
  0\\
   0\\
   0\\
   0\\
  -1.4922 +21.7322i\\
   1.6156 -23.5296i\\
   0\\
   1.9057 -27.7531i\\
  -0.4553 + 6.6311i\\
  \end{array}
\right)$ &
$2.7734720648$ & $ 10^{-10}$ & $4$\\\hline
$\left(
  \begin{array}{c}
   3.1100 - 6.7355i\\
  -3.8978 + 0.9624i\\
  \end{array}
\right)$ &
$\left(
  \begin{array}{c}
     0\\
  -2.8400 - 6.8559i\\
   0.3460 + 0.8353i\\
  -1.0251 - 2.4746i\\
   0\\
   0\\
  -0.9302 - 2.2454i\\
   0\\
   0\\
  \end{array}
\right)$ &
$2.3345140250$ & $ 10^{-12}$ & $5$\\\hline
$\left(
  \begin{array}{c}
  -1.9233 + 0.0000i\\
   1.0429 - 0.4627i\\
   0.2625 + 0.3788i\\
  -0.3661 + 0.0000i\\
  \end{array}
\right)$ &
$\left(
  \begin{array}{c}
  0\\
   0\\
   0\\
   0\\
  -1.4005 - 0.0000i\\
   0.8738 + 0.0000i\\
   0\\
  -0.0652 + 0.0000i\\
  -0.2482 - 0.0000i\\
  \end{array}
\right)$ &
$0.6052190547$ & $ 10^{-10}$ & $6$\\
\hline \end{tabular}
\end{scriptsize}
\end{table}
\begin{table}[t]
\centering 
\begin{scriptsize}
\begin{tabular}{|c|c|c|c|c|} \hline
$ F^{a_{2}a_{1}}$ & $B_{a_2}(u_{2})B_{a_{1}}(u_1)F^{a_{2}a_{1}}\ket{\Psi_0}$ & $E_{n}$ & $H\ket{\Psi_n}-E_n\ket{\Psi_n}$ & $n$\\\hline
$\left(
  \begin{array}{c}
  2.2480 +11.0915i\\
  -2.2334 -26.9793i\\
  \end{array}
\right)$ &
$\left(
  \begin{array}{c}
    0\\
  -0.2600 - 0.3755i\\
   0.8287 + 1.1967i\\
   0.7225 + 1.0434i\\
   0\\
   0\\
  -1.6224 - 2.3429i\\
   0\\
   0\\
  \end{array}
\right)$ & $-0.1238250060$ & $10^{-12}$ & $7$\\\hline
$\left(
  \begin{array}{c}
-0.2659 - 0.2496i\\
   0.5958 + 0.6008i\\
   0.1405 + 0.1120i\\
  -0.2335 - 0.2782i\\
  \end{array}
\right)$ &
$\left(
  \begin{array}{c}
   0\\
   0\\
   0\\
   0\\
   0.6301 + 0.3129i\\
   2.7450 + 1.3631i\\
   0\\
  -2.0208 - 1.0035i\\
  -0.6605 - 0.3280i\\
  \end{array}
\right)$ & $-1.6800801819$ & $10^{-10}$ & $8$\\\hline
$\left(
  \begin{array}{c}
   7.5081 - 1.5895i\\
  -3.1063 - 0.1958i\\
  \end{array}
\right)$ &
$\left(
  \begin{array}{c}
  0\\
  -1.3958 + 1.0506i\\
   0.2580 - 0.1942i\\
   2.9832 - 2.2454i\\
   0\\
   0\\
  -1.1488 + 0.8647i\\
   0\\
   0\\
  \end{array}
\right)$ & $-2.5834417305$ & $ 10^{-12}$ & $9$\\\hline
\end{tabular}
\end{scriptsize}
\caption{\label{table3}
Eigenstate \eqref{bethe state} with vector components $\{\mathcal{F}^{a_{M} a_{M-1} \dots a_{1}}\}$ given by \eqref{bethe state2}
where $N=2$ with the parameters $\eta=0.2, \zeta=0.1, c=1, c_{1}=-0.3,
\zeta'=-0.4, c'=-0.3, c_{1}'=-0.7, \alpha_-=0.8940, \alpha_+= 0.2704, \beta_-=-0.0321-1.5708i, \beta_+=0.8573, \theta_-=-0.0779-1.5708i, \theta_+=0.2604+3.1416i, p=1.5418i$ and $q=0.9267$ for the case of $E_{n}$ is the corresponding eigenenergy.
}
\end{table}

Now we are ready to write
out the eigenvalues $\Lambda(u)$ of the transfer matrices $t(u)$ as
\begin{eqnarray}
\Lambda(u)
&=&\frac{\sinh(2u+3\eta)}{\sinh(2u+\eta)} K^{(1)}(u)a_0(u)
\frac{Q^{(1)}(u-\eta)}{Q^{(1)}(u)}
\nonumber \\[4pt]
&&\,+ \frac{\sinh
2u\sinh(2u+3\eta)}{\sinh(2u+\eta)\sinh(2u+2\eta)}K^{(2)}(u+\frac{1}{2}\eta)b_0(u)\frac{Q^{(1)}(u+\eta)Q^{(2)}(u-\frac{1}{2}\eta)}
{Q^{(1)}(u)Q^{(2)}(u+\frac{1}{2}\eta)}
\nonumber \\[4pt]
&&\, + \frac{\sinh 2u}{\sinh(2u+2\eta)} K^{(3)}(u+\frac{1}{2}\eta)b_0(u)
\frac{Q^{(2)}(u+\frac{3}{2}\eta)}{Q^{(2)}(u+\frac{1}{2}\eta)}\no\\
&&\, +\,\sinh 2u\sinh(2u+3\eta)b_0(u)
\frac{hQ^{(1)}(u+\eta)}{Q^{(2)}(u+\frac{1}{2}\eta)},\label{T-Q relation}
\end{eqnarray}
\noindent
where the $2M$ Bethe roots must satisfy the BAEs
\eqref{BAES01} and \eqref{BAES02}, namely
\be
1&+&\frac{\sinh(2g_l+{3}\eta)}{\sinh(2g_l+\eta)}\frac{K^{(2)}(g_l+\frac{1}{2}\eta)}{K^{(3)}(g_l+\frac{1}{2}\eta)}
\frac{Q^{(1)}(g_l+\eta)Q^{(2)}(g_l-\frac{1}{2}\eta)}{Q^{(1)}(g_l)Q^{(2)}(g_l+\frac{3}{2}\eta)}\no\\[6pt]
&=&-h\,{\sinh(2g_l+2\eta)\sinh(2g_l+3\eta)}
\frac{Q^{(1)}(g_l+\eta)}
{K^{(3)}(g_l+\frac{1}{2}\eta)Q^{(2)}(g_l+\frac{3}{2}\eta)},\nonumber\\
&& l=1,\ldots, M, \label{BAES1}\\
-1&=&\frac{\sinh
(2u_k)}{\sinh(2u_k+2\eta)}\frac{K^{(2)}({u_k}+\frac{1}{2}\eta)b_0(u_k)}{K^{(1)}({u_k})a_0(u_k)}
\frac{Q^{(1)}(u_k+\eta)Q^{(2)}(u_k-\frac{1}{2}\eta)}{Q^{(1)}(u_k-\eta)Q^{(2)}(u_k+\frac{1}{2}\eta)},\nonumber\\
 && k=1,\ldots M.\label{BAES2}
\ee

The eigenvalue the Hamiltonian \eqref{hamilton} can be obtained by
\begin{eqnarray}
&&E=\sinh\eta
\frac{\partial \ln \Lambda(u)}{\partial u}|_{u=0,\{\theta_j\}=0}.\label{energy}
\end{eqnarray}
Here we present the numerical results for the $N=2$ and $N=3$ in Table~\ref{table1} and
Table~\ref{table2} respectively. From these Tables, we can see that the eigenvalues obtained from \eqref{energy} are the same as that from the exact diagonalization of the Hamiltonian
\eqref{hamilton}. Moreover, numerical results also show that the energy spectrum are complete.\\
Now, the eigenstate of the system can be written out explicitly as
\begin{eqnarray}
\label{bethe state sun}
\Psi &=& \sum_{\{a_i=1\}}^2\langle a_1,a_2,\cdots,a_M|\mathcal{F}\rangle
 B_{a_{M}}(u_{M}) B_{a_{M-1}}(u_{M-1}) \dots B_{a_{1}}(u_{1})\ket{\Psi_0}\nonumber\\
 &=&\sum_{\{a_i=1\}}^2 \mathcal{F}^{a_1 a_2 \cdots a_M} B_{a_{M}}(u_{M}) B_{a_{M-1}}(u_{M-1}) \dots B_{a_{1}}(u_{1})\ket{\Psi_0}.\label{oo}
\end{eqnarray}
As we mentioned above, the $\mathcal{F}^{a_{1}\dots a_{M}}$ are
the vector components of the nested Bethe states ( For example, for $N=2,M=2$, the state can be expressed by $\mathcal{F}^{11} B_{1}(u_{2}) B_{1}(u_{1})+\mathcal{F}^{12} B_{2}(u_{2}) B_{1}(u_{1})+
\mathcal{F}^{21} B_{1}(u_{2}) B_{2}(u_{1})+\mathcal{F}^{22}B_{2}(u_{2}) B_{2}(u_{1})$
). The numerical solutions of
the states constructed by \eqref{bethe state} with vector components $\{\mathcal{F}^{a_{M} a_{M-1} \dots a_{1}}\}$
given by \eqref{bethe state2} are shown in Table~\ref{table3}, which give rise to the eigenstates of the system.
Lastly, we should note that both the Bethe state given by \eqref{oo}
and the eigenvalue $\Lambda(u)$ given by \eqref{T-Q relation} have well-defined homogeneous limit (i.e., $\theta_j\rightarrow 0$).
This implies that in the homogeneous limit, the resulting Bethe states and the
eigenvalues give rise to the eigenstates and the corresponding eigenvalues of the model described by the Hamiltonian \eqref{hamilton}.


\section{Concluding remarks}
\label{sec:concluding remarks}

In this paper, we have studied the trigonometric $SU(3)$
model with the most generic integrable  boundary condition, which is described by the
Hamiltonian \eqref{hamilton} and the corresponding integrable boundary terms are
associated with  the most generic non-diagonal $K$-matrices given by
\eqref{Kmatrixfu}-\eqref{Kzheng matrix}. By combining the  algebraic Bethe ansatz and the
off-diagonal Bethe ansatz, we construct the eigenstates of the transfer matrix in terms of
the nested Bethe states given by \eqref{bethe state} and \eqref{bethe state2}, which have well-defined
homogeneous limit. The corresponding eigenvalues are given in terms of the inhomogeneous $T-Q$
relation \eqref{T-Q relation} and the associated BAEs \eqref{BAES1}-\eqref{BAES2}.

There are three forms of the off-diagonal reflection matrices $K^{\pm}$, i.e., the upper block form, the X form
and the lower block form \cite{Guangliang Li}. In this paper, we only consider the case that both the $K^{+}$ and the $K^{-}$ have a lower block form.
Our method can also be applied to the cases that both the $K^{+}$ and the $K^{-}$ have an upper block or X form  simultaneously:
For the upper $K^\pm$-matric block form, the eigenstate of the transfer matrix can be constructed as $
C_{a_{M}}(u_{M}) C_{a_{M-1}}(u_{M-1}) \dots C_{a_{1}}(u_{1})
\mathcal{F}^{a_{M} a_{M-1} \dots a_{1}} \ket{\Psi_0}$, where $\ket{\Psi_0}=( 0, 0, 1)^t$, $C_1=\mathbb{T}_{31}$, $C_2=\mathbb{T}_{32}$, and $\mathbb{T}_{ij}$ is the element of matrix $\mathbb{T}$ with the ith row and the jth column. While for the X form, the corresponding eigenstate can be constructed as $
X_{a_{M}}(u_{M}) X_{a_{M-1}}(u_{M-1}) \dots X_{a_{1}}(u_{1})
\mathcal{F}^{a_{M} a_{M-1} \dots a_{1}} \ket{\Psi_0}$, where $\ket{\Psi_0}=(0, 1, 0)^t$, $X_1=\mathbb{T}_{21}$ and $X_2=\mathbb{T}_{23}$.
We can prove that the eigenstate $\ket{\mathcal{F}}$ can be constructed by a similar nested Bethe states of an open spin-1/2 chain of length M with non-diagonal boundary terms. However,  for the case that the $K^-$ and $K^+$ have a different form, it seems that there does not exist an obvious reference state to which the first level algebraic Bethe ansatz can be applied, which leads to the eigenstate problem of the corresponding model being still  non-trivial.

The exact solution in this paper provides the basis for further analyzing the thermodynamic properties and the correlation functions of the model. In particular, the explicit expressions (\ref{bethe state sun}) and (\ref{bethe state2}) enable one to calculate the scalar products of Bethe states, by which the correlation behavior of some local operators can be computed further. Moreover, the $T-Q$ relation and the associated BAEs allow one to calculate the boundary energy of the model in the thermodynamic limit.

\section*{Acknowledgments}
We would like to thank Prof. Y. Wang for his valuable discussions and continuous encouragements.
The financial supports from the National Program
for Basic Research of MOST (Grant No. 2016YFA0300600 and
2016YFA0302104), the National Natural Science Foundation of China
(Grant Nos. 11434013, 11425522, 11547045, 11774397, 11775178 and 11775177), the Major Basic Research Program of Natural Science of Shaanxi Province
(Grant Nos. 2017KCT-12, 2017ZDJC-32) and the Strategic Priority Research Program of the Chinese
Academy of Sciences are gratefully acknowledged.
P. Sun and F. Wen are also partially supported by the NWU graduate student innovation funds No. YZZ15088 and No. YYB17003.



\begin{thebibliography}{99}
\bibitem{1cn Yang}
C.-N. Yang, Phys. Rev. Lett. {\bf 19} (1967) 1312; \\
C.-N. Yang, Phys. Rev. {\bf 168} (1968) 1920.
\bibitem{2RJ Baxter}
R.~J. Baxter, Phys. Rev. Lett. {\bf 26} (1971) 832;\\
R.~J. Baxter, Phys. Rev. Lett. {\bf 26} (1971) 834;\\
R.~J. Baxter, Ann. Phys.(N.Y.) {\bf 70} (1972) 323.
\bibitem{3Baxter 2}
R.~J. Baxter, {\it Exactly Solved Models in Statistical Mechanics}, Academic Press, 1982.
\bibitem{4CT Black}
C.~T. Black, D.~C. Ralph and M.~Tinkham, Phys. Rev. Lett. {\bf 76} (1996) 688;\\
C.~T. Black, D.~C. Ralph and M.~Tinkham, Phys. Rev. Lett. {\bf 78} (1997) 4087.
\bibitem{5J Dukelsky}
J.~Dukelsky, S.~Pittel and G.~Sierra,  Rev. Mod. Phys. {\bf 76} (2004) 643.
\bibitem{6BM Mccoy}
B.~M. McCoy, {\it Advanced Statistical Mechanics}, Oxford University Press, 2010.
\bibitem{7T Giamarchi}
T.~Giamarchi, {\it Quantum Physics in One Dimension}, Oxford University Press, 2003.
\bibitem{8V Chari}
V.~Chari and A.~Pressley, {\it A Guide to Quantum Groups}, Cambridge University Press, 1994.
\bibitem{10L.A. Takhtadzhan}
L.~A. Takhtadzhan and L.~D. Faddeev, Rush. Math. Surveys {\bf 34} (1979) 11.
\bibitem{11E.K. Sklyanin}
E.~K. Sklyanin and L.~D. Faddeev, Sov. Phys. Dokl. {\bf 23} (1978) 902.
\bibitem{12F.C. Alcaraz}
F.~C. Alcaraz, M.~N. Barber, M.~T. Batchelor, R.~J. Baxter and G.~R.~W. Quispel, J. Phys.A {\bf 20} (1987) 6397.
\bibitem{13E.K. Sklyanin}
E.~K. Sklyanin, J. Phys. A {\bf 21} (1988) 2375.
\bibitem{14J.Cao1}
J.~Cao, W.-L. Yang, K.~Shi and Y.~Wang, Phys. Rev. Lett. {\bf 111} (2013) 137201.
\bibitem{15J.Cao2}
J.~Cao, W.-L. Yang, K.~Shi and Y.~Wang, Nucl. Phys. B {\bf 875} (2013) 152.
\bibitem{16J.Cao3}
J.~Cao, S.~Cui, W.-L. Yang, K.~Shi and Y.~Wang, Nucl. Phys. B {\bf 886} (2014) 185.
\bibitem{17J.Cao4}
J.~Cao, W.-L. Yang, K.~Shi and Y.~Wang, Nucl. Phys. B {\bf 877} (2013) 152.
\bibitem{18J.Cao5}
J.~Cao, W.-L. Yang, K.~Shi and Y.~Wang, J. High Energy Phys. {\bf 04} (2014) 143.
\bibitem{19J.Cao6}
Y.-Y. Li, J.~Cao, W.-L. Yang, K.~Shi and Y.~Wang, Nucl. Phys. B {\bf 884} (2014) 17.
\bibitem{20X.Zhang}
X.~Zhang, J.~Cao, W.-L. Yang, K.~Shi and Y.~Wang, J. Stat. Mech. (2014) P04031.
\bibitem{21K.Hao}
K.~Hao, J.~Cao, G.-L. Li, W.-L. Yang, K.~Shi and Y.~Wang, J. High Energy Phys. {\bf 06} (2014) 128
\bibitem{22Y.Wang}
Y.~Wang, W.-L. Yang, J.~Cao and K.~Shi, {\it Off-Diagonal Bethe Ansatz for Exactly Solvable Models}, Springer Press, 2015.

\bibitem{jia Zhang state1}
X.~Zhang, Y.-Y. Li, J.~Cao, W.-L. Yang, K.~Shi and Y.~Wang, Nucl. Phys. B {\bf 893} (2015) 70.

\bibitem{jia zhang state2}
X.~Zhang, Y.-Y. Li, J.~Cao, W.-L. Yang, K.~Shi and Y.~Wang, J. Stat. Mech. (2015) P05014.

\bibitem{jia Niccoli1}
G.~Niccoli, J. Stat. Mech. (2012) P10025.

\bibitem{jia Niccoli2}
G.~Niccoli, Nucl. Phys. B {\bf 870} (2013) 397;\\
G.~Niccoli, J. Phys. A: Math. Theor. {\bf 46} (2013) 075003.


\bibitem{jia Niccoli3}
S.~Faldella, N.~Kitanine and G.~Niccoli, J. Stat. Mech. (2014) P01011.


\bibitem{1}
S. Belliard, Nucl. Phys. B {\bf 892} (2015) 1.
\bibitem{2}
S. Belliard and R.A. Pimenta, Nucl. Phys. B {\bf 894} (2015) 527.
\bibitem{3}
J. Avan, S. Belliard, N. Grosjean and R. A. Pimenta, Nucl. Phys. B {\bf 899} (2015) 229.


\bibitem{Guangliang Li}
G.-L. Li, J.~Cao, K.~Hao, F.-K. Wen, W.-L. Yang and  K.~Shi, Nucl. Phys. B {\bf 910} (2016) 410.

\bibitem{361J.H.H. Perk}
J.~H.~H. Perk and C.~L. Schultz, Phys. Lett. A {\bf 84} (1981) 407


\bibitem{36J.H.H. Perk}
J.~H.~H. Perk and C.~L. Schultz, {\it Families of commuting transfer matrices in q-state vertex models, in: M.~Jimbo, T.~Miwa (Eds.), Non-Linear Integrable Systems-Classical Theory and Quantum Theory}, World Scientific, 1983, pp. 135-152.

\bibitem{37C.L. Schultz}
C.~L. Schultz, Physica A {\bf 122} (1983) 71.
\bibitem{38J.H.H. Perk}
J.~H.~H. Perk and H. Au-Yang, {\it Yang-Baxter Equation, in: J.-P. Fran\c{c}oise, G.L. Naber, T.S. Tsun (Eds.), Encyclopedia of Mathematical Physics}, Academic Press, 2006, pp. 465-473
\bibitem{39V.V. Bazhanov}
V.~V. Bazhanov, Phys. Lett. B {\bf 159} (1985) 321.
\bibitem{40M.Jimbo}
M.~Jimbo, Commun. Math. Phys. {\bf 102} (1986) 537.

\end{thebibliography}
\end{document}